\pdfoutput=1

\documentclass[12pt,a4paper]{article}

\usepackage{ifthen} 
\newboolean{pdflatex}
\setboolean{pdflatex}{true} 

\newboolean{articletitles}
\setboolean{articletitles}{true} 

\newboolean{uprightparticles}
\setboolean{uprightparticles}{false} 

\newboolean{inbibliography}
\setboolean{inbibliography}{false} 

\def\paperauthors{LHCb collaboration} 
\def\paperasciititle{Measurement of the lifetime of the doubly
  charmed baryon Xicc++} 
\def\papertitle{Measurement of the lifetime of the doubly
  charmed baryon $\Xiccpp$} 
\def\paperkeywords{{High Energy Physics}, {LHCb}} 
\def\papercopyright{\the\year\ CERN for the benefit of the LHCb collaboration} 
\def\paperlicence{CC-BY-4.0 licence}
\def\paperlicenceurl{https://creativecommons.org/licenses/by/4.0/}

\usepackage[top=1in, bottom=1.25in, left=1in, right=1in]{geometry}

\usepackage{microtype}
\usepackage{lineno}  
\usepackage{xspace} 
\usepackage{caption} 

\usepackage{graphicx}  
\usepackage{color}
\usepackage{colortbl}
\graphicspath{{./figs/}} 

\usepackage{amsmath} 
\usepackage{amssymb}
\usepackage{amsfonts}
\usepackage{upgreek} 

\newcommand*\patchAmsMathEnvironmentForLineno[1]{%
\expandafter\let\csname old#1\expandafter\endcsname\csname #1\endcsname
\expandafter\let\csname oldend#1\expandafter\endcsname\csname
end#1\endcsname
 \renewenvironment{#1}%
   {\linenomath\csname old#1\endcsname}%
   {\csname oldend#1\endcsname\endlinenomath}%
}
\newcommand*\patchBothAmsMathEnvironmentsForLineno[1]{%
  \patchAmsMathEnvironmentForLineno{#1}%
  \patchAmsMathEnvironmentForLineno{#1*}%
}
\AtBeginDocument{%
\patchBothAmsMathEnvironmentsForLineno{equation}%
\patchBothAmsMathEnvironmentsForLineno{align}%
\patchBothAmsMathEnvironmentsForLineno{flalign}%
\patchBothAmsMathEnvironmentsForLineno{alignat}%
\patchBothAmsMathEnvironmentsForLineno{gather}%
\patchBothAmsMathEnvironmentsForLineno{multline}%
\patchBothAmsMathEnvironmentsForLineno{eqnarray}%
}

\usepackage{hyperxmp}

\usepackage[pdftex,
            pdfauthor={\paperauthors},
            pdftitle={\paperasciititle},
            pdfkeywords={\paperkeywords},
            pdfcopyright={Copyright (C) \papercopyright},
            pdflicenseurl={\paperlicenceurl}]{hyperref}

\usepackage[all]{hypcap} 

\usepackage{xspace} 
\usepackage{upgreek}

\def\lhcb {\mbox{LHCb}\xspace}

\def\babar  {\mbox{BaBar}\xspace}
\def\belle  {\mbox{Belle}\xspace}

\def\MagUp {\mbox{\em Mag\kern -0.05em Up}\xspace}

\ifthenelse{\boolean{uprightparticles}}%
{

 \def\Ppi         {\ensuremath{\uppi}\xspace}

 \def\PDelta      {\ensuremath{\Delta}\xspace}                 
 \def\PXi      {\ensuremath{\Xi}\xspace}                 
 \def\PLambda      {\ensuremath{\Lambda}\xspace}                 
 \def\PSigma      {\ensuremath{\Sigma}\xspace}                 
 \def\POmega      {\ensuremath{\Omega}\xspace}                 
 \def\PUpsilon      {\ensuremath{\Upsilon}\xspace}                 
 
 \def\PB      {\ensuremath{\mathrm{B}}\xspace}                 
                  
 \def\PD      {\ensuremath{\mathrm{D}}\xspace}

 \def\PK      {\ensuremath{\mathrm{K}}\xspace}

 \def\Pb      {\ensuremath{\mathrm{b}}\xspace}                 
 \def\Pc      {\ensuremath{\mathrm{c}}\xspace}

 \def\Pi      {\ensuremath{\mathrm{i}}\xspace}

 \def\Pp      {\ensuremath{\mathrm{p}}\xspace}

}
{

 \def\Ppi         {\ensuremath{\pi}\xspace}

 \mathchardef\PDelta="7101
 \mathchardef\PXi="7104
 \mathchardef\PLambda="7103
 \mathchardef\PSigma="7106
 \mathchardef\POmega="710A
 \mathchardef\PUpsilon="7107
                  
 \def\PB      {\ensuremath{B}\xspace}                 
                  
 \def\PD      {\ensuremath{D}\xspace}

 \def\PK      {\ensuremath{K}\xspace}

 \def\Pb      {\ensuremath{b}\xspace}                 
 \def\Pc      {\ensuremath{c}\xspace}

 \def\Pi      {\ensuremath{i}\xspace}

 \def\Pp      {\ensuremath{p}\xspace}

}

\makeatletter
\ifcase \@ptsize \relax
  \newcommand{\miniscule}{\@setfontsize\miniscule{4}{5}}
\or
  \newcommand{\miniscule}{\@setfontsize\miniscule{5}{6}}
\or
  \newcommand{\miniscule}{\@setfontsize\miniscule{5}{6}}
\fi
\makeatother

\DeclareRobustCommand{\optbar}[1]{\shortstack{{\miniscule (\rule[.5ex]{1.25em}{.18mm})}
  \\ [-.7ex] $#1$}}

\def\cquark    {{\ensuremath{\Pc}}\xspace}

\def\bquark    {{\ensuremath{\Pb}}\xspace}

\def\pion   {{\ensuremath{\Ppi}}\xspace}

\def\pip    {{\ensuremath{\pion^+}}\xspace}
\def\pim    {{\ensuremath{\pion^-}}\xspace}

\def\kaon    {{\ensuremath{\PK}}\xspace}
  \def\Kbar    {{\kern 0.2em\overline{\kern -0.2em \PK}{}}\xspace}

\def\KorKbar    {\kern 0.18em\optbar{\kern -0.18em K}{}\xspace}

\def\Km      {{\ensuremath{\kaon^-}}\xspace}

  \def\Dbar    {{\kern 0.2em\overline{\kern -0.2em \PD}{}}\xspace}
\def\D       {{\ensuremath{\PD}}\xspace}

\def\DorDbar    {\kern 0.18em\optbar{\kern -0.18em D}{}\xspace}

\def\Dp      {{\ensuremath{\D^+}}\xspace}

\def\Bbar    {{\ensuremath{\kern 0.18em\overline{\kern -0.18em \PB}{}}}\xspace}

\def\BorBbar    {\kern 0.18em\optbar{\kern -0.18em B}{}\xspace}

  \def\Y#1S{\ensuremath{\PUpsilon{(#1S)}}\xspace}

\def\proton      {{\ensuremath{\Pp}}\xspace}

\def\Xires       {{\ensuremath{\PXi}}\xspace}
\def\Xiresbar    {{\ensuremath{\overline \Xires}}\xspace}
\def\Lz          {{\ensuremath{\PLambda}}\xspace}
\def\Lbar        {{\ensuremath{\kern 0.1em\overline{\kern -0.1em\PLambda}}}\xspace}
\def\LorLbar    {\kern 0.18em\optbar{\kern -0.18em \PLambda}{}\xspace}

\def\Lb      {{\ensuremath{\Lz^0_\bquark}}\xspace}

\def\Lc      {{\ensuremath{\Lz^+_\cquark}}\xspace}

\def\to                 {\ensuremath{\rightarrow}\xspace}

\def\AT#1     {\ensuremath{A_{\mathrm{T}}^{#1}}\xspace}           

\def\C#1      {\ensuremath{\mathcal{C}_{#1}}\xspace}                       
\def\Cp#1     {\ensuremath{\mathcal{C}_{#1}^{'}}\xspace}                    
\def\Ceff#1   {\ensuremath{\mathcal{C}_{#1}^{\mathrm{(eff)}}}\xspace}        
\def\Cpeff#1  {\ensuremath{\mathcal{C}_{#1}^{'\mathrm{(eff)}}}\xspace}       
\def\Ope#1    {\ensuremath{\mathcal{O}_{#1}}\xspace}                       
\def\Opep#1   {\ensuremath{\mathcal{O}_{#1}^{'}}\xspace}                    

\newcommand{\tev}{\ifthenelse{\boolean{inbibliography}}{\ensuremath{~T\kern -0.05em eV}}{\ensuremath{\mathrm{\,Te\kern -0.1em V}}}\xspace}
\newcommand{\gev}{\ensuremath{\mathrm{\,Ge\kern -0.1em V}}\xspace}
\newcommand{\mev}{\ensuremath{\mathrm{\,Me\kern -0.1em V}}\xspace}
\newcommand{\kev}{\ensuremath{\mathrm{\,ke\kern -0.1em V}}\xspace}
\newcommand{\ev}{\ensuremath{\mathrm{\,e\kern -0.1em V}}\xspace}
\newcommand{\gevc}{\ensuremath{{\mathrm{\,Ge\kern -0.1em V\!/}c}}\xspace}
\newcommand{\mevc}{\ensuremath{{\mathrm{\,Me\kern -0.1em V\!/}c}}\xspace}
\newcommand{\gevcc}{\ensuremath{{\mathrm{\,Ge\kern -0.1em V\!/}c^2}}\xspace}
\newcommand{\gevgevcccc}{\ensuremath{{\mathrm{\,Ge\kern -0.1em V^2\!/}c^4}}\xspace}
\newcommand{\mevcc}{\ensuremath{{\mathrm{\,Me\kern -0.1em V\!/}c^2}}\xspace}

\def\invfb   {\ensuremath{\mbox{\,fb}^{-1}}\xspace}

\def\ps   {\ensuremath{{\mathrm{ \,ps}}}\xspace}
\def\fs   {\ensuremath{\mathrm{ \,fs}}\xspace}

\def\gsim{{~\raise.15em\hbox{$>$}\kern-.85em
          \lower.35em\hbox{$\sim$}~}\xspace}
\def\lsim{{~\raise.15em\hbox{$<$}\kern-.85em
          \lower.35em\hbox{$\sim$}~}\xspace}

\def\sPlot{\mbox{\em sPlot}\xspace}

\def\evtgen     {\mbox{\textsc{EvtGen}}\xspace}

\def\geant      {\mbox{\textsc{Geant4}}\xspace}

\def\photos     {\mbox{\textsc{Photos}}\xspace}

\def\pythia     {\mbox{\textsc{Pythia}}\xspace}

\def\tell1  {TELL1\xspace}
\def\ukl1   {UKL1\xspace}

\newcommand{\eg}{\mbox{\itshape e.g.}\xspace}
\newcommand{\ie}{\mbox{\itshape i.e.}\xspace}

\def\Lcp          {{\ensuremath{\Lz^+_\cquark}}\xspace}

\def\Xiccbare     {{\ensuremath{\Xires_{\cquark\cquark}}}\xspace}
\def\Xiccpp       {{\ensuremath{\Xires^{++}_{\cquark\cquark}}}\xspace}

\def\Xiccp        {{\ensuremath{\Xires^{+}_{\cquark\cquark}}}\xspace}

\usepackage{cite} 
\usepackage{mciteplus}

\usepackage{longtable} 

\usepackage{xspace} 
\usepackage{upgreek}

\def\DeltaOrDeltabar  {\kern 0.18em\optbar{\kern -0.18em \PDelta}{}\xspace}
\def\XiOrXibar        {\kern 0.18em\optbar{\kern -0.18em \PXi}{}\xspace}
\def\SigmaOrSigmabar  {\kern 0.18em\optbar{\kern -0.18em \PSigma}{}\xspace}
\def\OmegaOrOmegabar  {\kern 0.18em\optbar{\kern -0.18em \POmega}{}\xspace}

\def\Lcp          {{\ensuremath{\Lz^+_\cquark}}\xspace}

\def\Xiccbare     {{\ensuremath{\Xires_{\cquark\cquark}}}\xspace}
\def\Xiccpp       {{\ensuremath{\Xires^{++}_{\cquark\cquark}}}\xspace}

\def\Xiccp        {{\ensuremath{\Xires^{+}_{\cquark\cquark}}}\xspace}

\newcommand{\TeVnosp}{\ifthenelse{\boolean{inbibliography}}{\ensuremath{~T\kern -0.05em eV}}{\ensuremath{\mathrm{Te\kern -0.1em V}}}}
\newcommand{\GeVnosp}{\ensuremath{\mathrm{Ge\kern -0.1em V}}}
\newcommand{\MeVnosp}{\ensuremath{\mathrm{Me\kern -0.1em V}}}
\newcommand{\keVnosp}{\ensuremath{\mathrm{ke\kern -0.1em V}}}
\newcommand{\eVnosp}{\ensuremath{\mathrm{e\kern -0.1em V}}}
\newcommand{\GeVcnosp}{\ensuremath{{\mathrm{Ge\kern -0.1em V\!/}c}}}
\newcommand{\MeVcnosp}{\ensuremath{{\mathrm{Me\kern -0.1em V\!/}c}}}
\newcommand{\GeVccnosp}{\ensuremath{{\mathrm{Ge\kern -0.1em V\!/}c^2}}}
\newcommand{\MeVccnosp}{\ensuremath{{\mathrm{Me\kern -0.1em V\!/}c^2}}}
\newcommand{\GeVGeVccccnosp}{\ensuremath{{\mathrm{Ge\kern -0.1em V^2\!/}c^4}}}

\def\genxicctwo   {\mbox{\textsc{GenXicc2.0}}\xspace}

\def\SymbolXiccTau{\ensuremath{\tau\left(\Xiccpp\right)}}
\def\SymbolLbTau{\ensuremath{\tau\left(\Lb\right)}}

\def\XiccTauStat{\ensuremath{0.256\,^{+0.024}_{-0.022} \ps}}
\def\XiccTau{\ensuremath{0.256\,^{+0.024}_{-0.022}\,{\rm(stat)\,}\pm
    0.014\,{\rm(syst)}\,\ps}}

\begin{document}

\renewcommand{\thefootnote}{\fnsymbol{footnote}}
\setcounter{footnote}{1}


\begin{titlepage}
\pagenumbering{roman}

\vspace*{-1.5cm}
\centerline{\large EUROPEAN ORGANIZATION FOR NUCLEAR RESEARCH (CERN)}
\vspace*{1.5cm}
\noindent
\begin{tabular*}{\linewidth}{lc@{\extracolsep{\fill}}r@{\extracolsep{0pt}}}
\ifthenelse{\boolean{pdflatex}}
{\vspace*{-1.5cm}\mbox{\!\!\!\includegraphics[width=.14\textwidth]{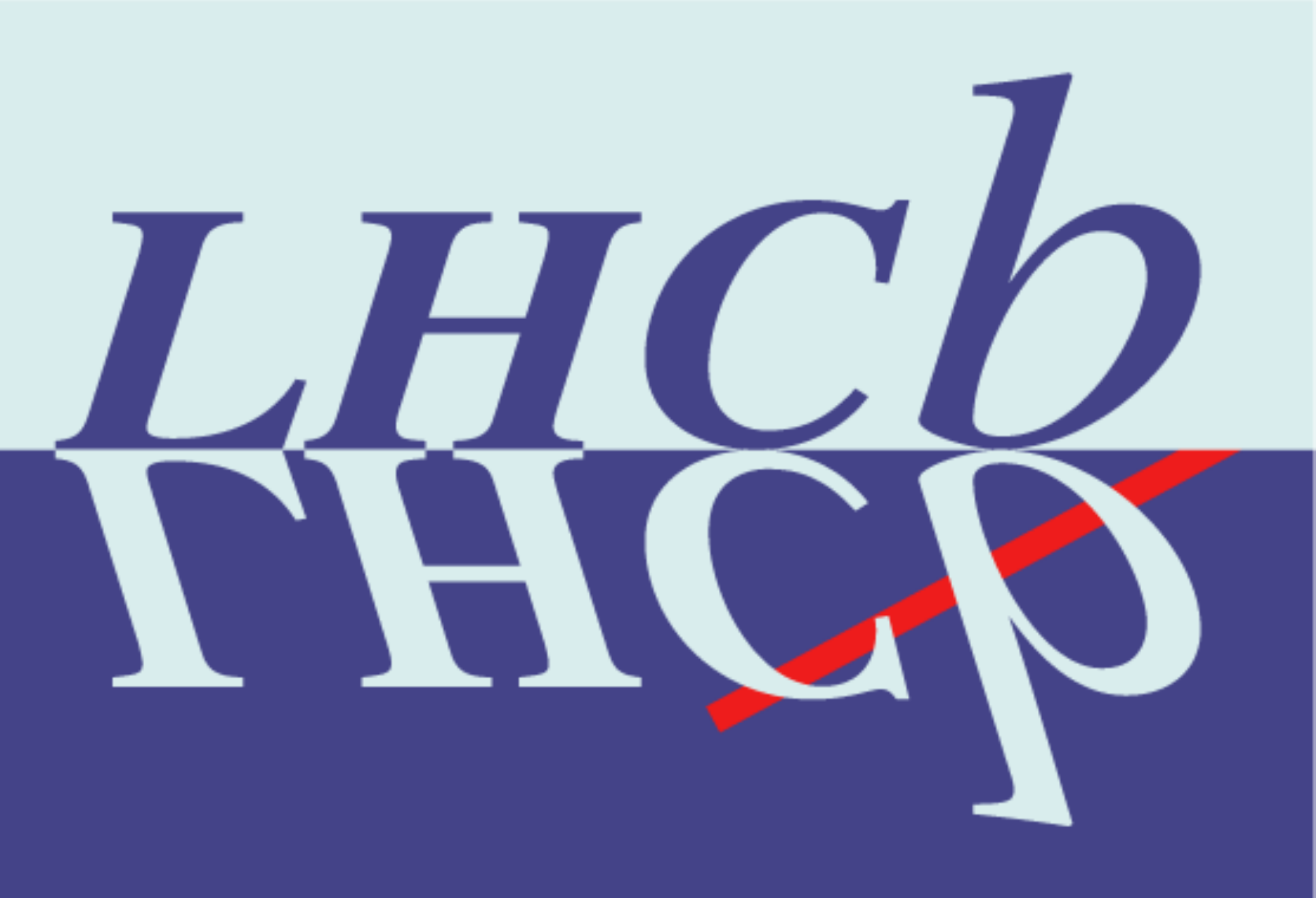}} & &}%
{\vspace*{-1.2cm}\mbox{\!\!\!\includegraphics[width=.12\textwidth]{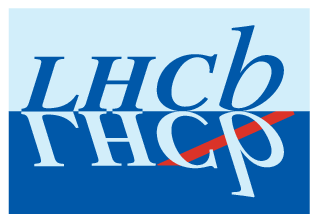}} & &}%
\\
 & & CERN-EP-2018-146 \\  
 & & LHCb-PAPER-2018-019 \\  
 & & July 31, 2018 \\ 
\end{tabular*}

\vspace*{4.0cm}

{\normalfont\bfseries\boldmath\huge
\begin{center}
  \papertitle 
\end{center}
}

\vspace*{2.0cm}

\begin{center}
\paperauthors\footnote{Authors are listed at the end of this Letter.}
\end{center}

\vspace{\fill}

\begin{abstract}
  \noindent
  The first measurement of the lifetime of the doubly charmed baryon
  $\Xiccpp$ is presented,
  with the signal reconstructed in the final state $\Lc \Km \pip
  \pip$.
  The data sample used corresponds to an integrated luminosity of
  $1.7\mbox{\,\rm fb}^{-1}$,
  collected by the LHCb experiment in
  proton-proton collisions at a centre-of-mass energy of $13\mathrm{\,Te\kern -0.1em V}$.
  The $\Xiccpp$ lifetime is measured to be 
  $0.256\,^{+0.024}_{-0.022}{\,\rm (stat)\,}  \pm 0.014
  {\,\rm(syst)}\mathrm{\,ps}$. 
\end{abstract}

\vspace*{2.0cm}

\begin{center}
  Published in Phys.~Rev.~Lett. 121, 052002 (2018)
\end{center}

\vspace{\fill}

{\footnotesize 
\centerline{\copyright~\papercopyright. \href{\paperlicenceurl}{\paperlicence}.}}
\vspace*{2mm}

\end{titlepage}


\newpage
\setcounter{page}{2}
\mbox{~}

\cleardoublepage


\renewcommand{\thefootnote}{\arabic{footnote}}
\setcounter{footnote}{0}



\pagestyle{plain} 
\setcounter{page}{1}
\pagenumbering{arabic}


%


The quark model of hadrons predicts the existence of
weakly decaying baryons that contain two beauty or charm quarks,
and are therefore referred to as doubly heavy baryons. Such states provide a unique
system for testing models of quantum chromodynamics (QCD),
the theory that describes the strong interaction.
In the quark model, the doubly charmed baryon \Xiccbare
forms an isodoublet, consisting of
the \Xiccpp and \Xiccp baryons 
with quark content $ccu$ and $ccd$, respectively. 
Predictions for the $\Xiccp$ lifetime span the range from 50 to
250\fs, 
while the \Xiccpp lifetime 
is expected to be three to four times larger, from 200 to
$1050\fs$~\cite{
Fleck:1989mb,
guberina1999inclusive,
Kiselev:1998sy,
Likhoded:1999yv,
Onishchenko:2000yp,
Anikeev:2001rk,
Kiselev:2001fw,
hsi2008lifetime,
Karliner:2014gca,
Berezhnoy:2016wix}. 
The predicted larger $\Xiccpp$ lifetime is due to 
the destructive Pauli interference of
the charm-quark decay products and the valence (up) quark in the initial state, 
whereas the $\Xiccp$ lifetime is shortened due to an additional 
contribution from $W$-exchange between the charm and down quarks~\cite{
Fleck:1989mb,
guberina1999inclusive,
Kiselev:1998sy,
Likhoded:1999yv,
Onishchenko:2000yp,
Anikeev:2001rk,
Kiselev:2001fw,
hsi2008lifetime,
Karliner:2014gca,
Berezhnoy:2016wix}.
Charge-conjugate processes are implied throughout this Letter. 

The SELEX collaboration~\cite{Mattson:2002vu,Ocherashvili:2004hi}
reported the observation of the $\Xiccp$ baryon 
in the final states $\Lcp \Km \pip$ and $\proton \Dp \Km$,
with a measured mass of $3518.7 \pm 1.7 \mevcc$.
Its lifetime was found to be less than $33\fs$ at the 90\% confidence level.
However, the signal
has not been confirmed in
searches performed at 
the FOCUS~\cite{Ratti:2003ez},
\babar~\cite{Aubert:2006qw}, 
\belle~\cite{Chistov:2006zj},
and \lhcb~\cite{LHCb-PAPER-2013-049} experiments. 
Recently, the LHCb collaboration
observed a resonance in the $\Lcp\Km\pip\pip$ mass spectrum at a mass of
$3621.40 \pm 0.78\mevcc$~\cite{LHCb-PAPER-2017-018}, which is
consistent with expectations for the $\Xiccpp$ baryon (\eg Ref.~\cite{Alexandrou:2017xwd}).
The difference in masses between the two reported states, $103 \pm 2 \mevcc$,
is much larger than the few\mevcc expected by 
the breaking of isospin
symmetry~\cite{Hwang:2008dj,Brodsky:2011zs,Karliner:2017gml}, 
and that is observed in all other isodoublets.
While the resonance seen in the $\Lcp\Km\pip\pip$ mass spectrum by LHCb is consistent with being the $\Xiccpp$ baryon,
a measurement of its lifetime is critical to establish its nature.
The lifetime is also a necessary ingredient for theoretical predictions of branching fractions
of  \Xiccbare decays, and can offer insight into the interplay between strong and weak interactions
in these decays.

This Letter reports the first measurement of the \Xiccpp lifetime,
with the \Xiccpp baryon reconstructed through the decay chain
$\Xiccpp \to \Lcp\Km\pip\pip$, $\Lcp \to p \Km \pip$.
The data sample used, the same as in Ref.~\cite{LHCb-PAPER-2017-018}, 
corresponds to an integrated luminosity of $1.7$\invfb,
collected by the LHCb experiment in
proton-proton collisions at a centre-of-mass energy of 13\tev.
Since the combined reconstruction and selection efficiency varies as a function of
the decay time, 
the decay-time distribution is measured relative to that of a control
mode with similar
topology and known lifetime~\cite{LHCb-PAPER-2014-003,PDG2017}, $\Lb \to \Lcp\pim\pip\pim$.
This technique, used in a number of lifetime measurements
at LHCb~\cite{LHCb-PAPER-2012-017,
LHCb-PAPER-2013-032,
LHCb-PAPER-2014-003,
LHCb-PAPER-2014-021, 
LHCb-PAPER-2014-037,
LHCb-PAPER-2014-048,
LHCb-PAPER-2014-060,
LHCb-PAPER-2016-008,
LHCb-PAPER-2017-004}, 
leads to a reduced systematic uncertainty as it is only sensitive 
to the ratio of the decay-time acceptances. 

The LHCb detector~\cite{Alves:2008zz,LHCb-DP-2014-002} is a
single-arm forward spectrometer covering the pseudorapidity range $2 < \eta < 5$, designed for
the study of particles containing \bquark\ or \cquark\ quarks. The detector elements that are particularly
relevant to this analysis are: a silicon-strip vertex detector~\cite{LHCb-DP-2014-001} surrounding the $pp$ interaction
region that allows \cquark\ and \bquark\ hadrons to be identified from their characteristically long
flight distance; a tracking system~\cite{LHCb-DP-2013-003},
placed upstream and downstream of a dipole magnet,
that provides a measurement of momentum, $p$, of charged
particles; and two ring-imaging Cherenkov detectors~\cite{LHCb-DP-2012-003} that are able to discriminate between
different species of charged hadrons.
The magnetic field polarity can be reverted periodically throughout the data-taking.
The online event-selection is performed by a trigger~\cite{LHCb-DP-2012-004}, 
which consists of a hardware stage, based on information from the calorimeter and muon
systems~\cite{LHCb-DP-2013-004, LHCb-DP-2013-001}, followed by a software stage, which applies a full event
reconstruction incorporating near-real-time
alignment and calibration of the
detector~\cite{LHCb-PROC-2015-011}.
The output of the reconstruction performed in the software trigger~\cite{LHCb-DP-2016-001}
is used as input to the present analysis.

Samples of simulated $pp$ collisions are generated using
\pythia~\cite{Sjostrand:2007gs,*Sjostrand:2006za} 
with a specific \lhcb
configuration~\cite{LHCb-PROC-2010-056}.  
A dedicated generator, \genxicctwo~\cite{Chang:2007pp,*Chang:2009va}, 
is used to simulate the production of the \Xiccpp baryon.
Decays of hadrons
are described by \evtgen~\cite{Lange:2001uf}, in which final-state
radiation is simulated using \photos~\cite{Golonka:2005pn}. The
interaction of the generated particles with the detector, and its response,
are implemented using the \geant
toolkit~\cite{Allison:2006ve, *Agostinelli:2002hh} as described in
Ref.~\cite{LHCb-PROC-2011-006}.

Candidate $\Xiccpp \to \Lcp\Km\pip\pip$ decays are reconstructed
and selected with a multivariate selector following the same procedure as
used in the previous analysis~\cite{LHCb-PAPER-2017-018},
except for two additional selection criteria.
The first requires that the events are selected, at the
hardware-trigger level, either by large transverse energy deposits in
the calorimeter from the decay products of the $\Xiccpp$ candidate or by
activity in the calorimeter or muon system from particles other than
the $\Xiccpp$ decay products. This requirement removes events for which the
efficiency can not be determined precisely.
The second is a requirement on the reconstructed decay time of the
$\Xiccpp$ candidates, $t$, 
which must lie in the range $0.1$--$2.0$\ps, where 
the lower limit on $t$ is imposed to avoid biases from resolution effects.
Candidate $\Lb \to \Lcp\pim\pip\pim$ decays are reconstructed
and selected in exactly the same way as $\Xiccpp$ decays,
except that the
allowed invariant-mass range is centred around the \Lb mass and both
negatively charged $\Lb$ decay products are required to be identified as pions.
The same hardware and software trigger criteria are applied
to both \Xiccpp and \Lb candidates.

\begin{figure}
\begin{center}
    \includegraphics*[width=.49\textwidth]{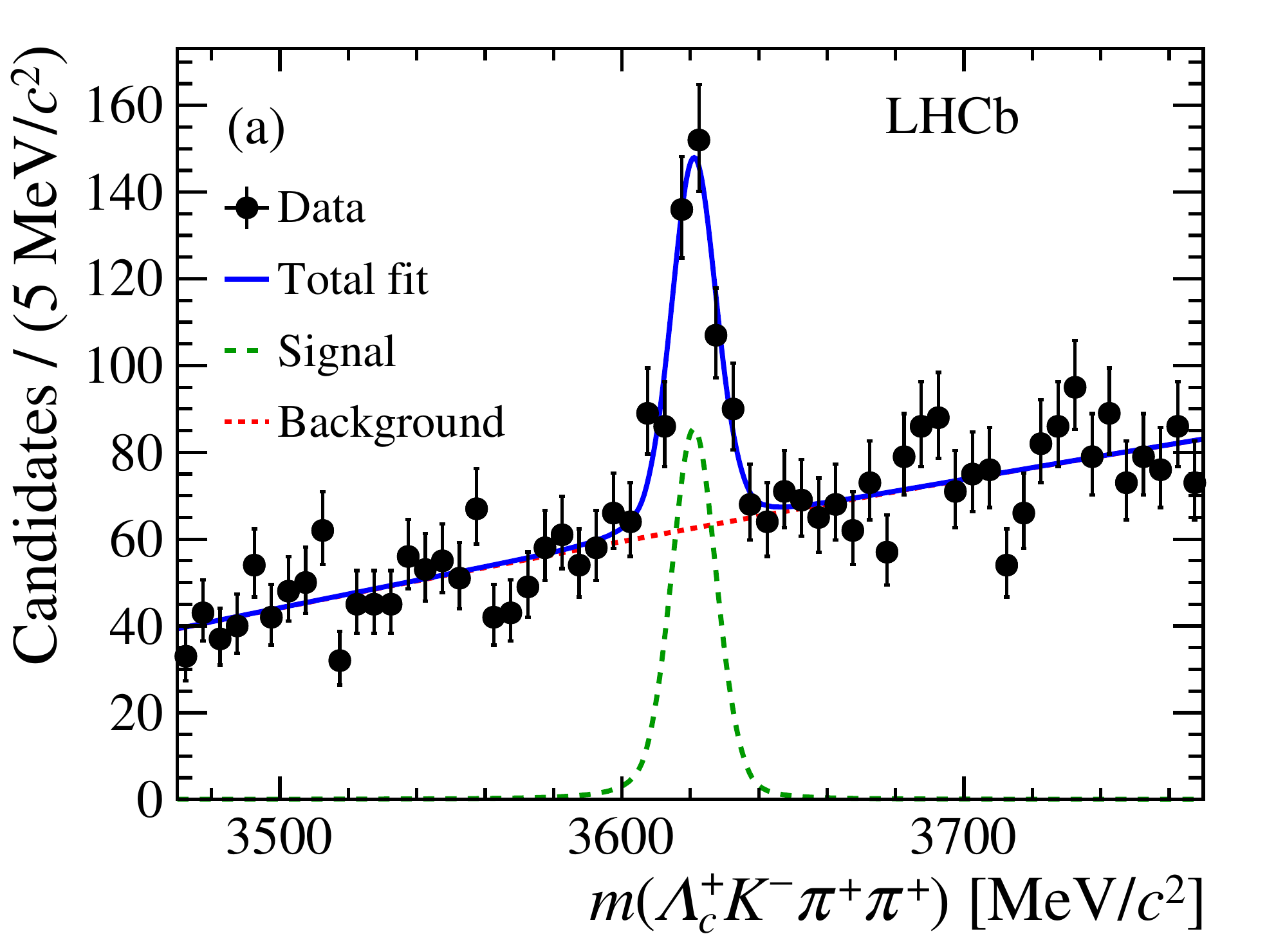}
    \includegraphics*[width=.49\textwidth]{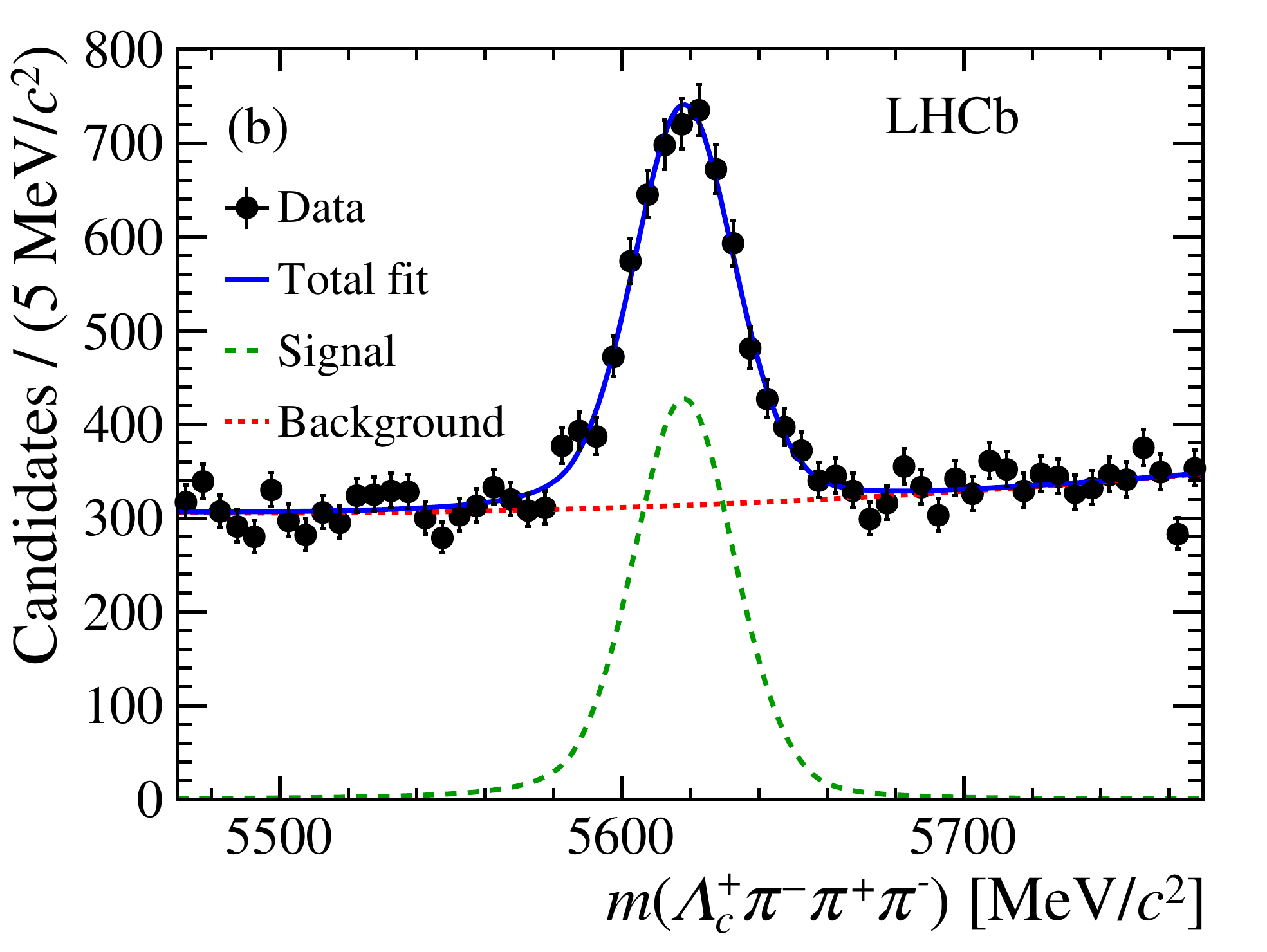}
\end{center}
\vspace*{-.7cm}
\caption{\small
  Invariant-mass distributions of (a) $\Xiccpp \to \Lcp\Km\pip\pip$ 
  and (b) $\Lb \to \Lcp \pim \pip\pim$ candidates, with fit results 
  shown.
}
\label{fig:MassSys-Xicc}
\end{figure}

To obtain better resolution, the invariant mass of a candidate is calculated as 
\begin{eqnarray}
        m &=& M(\Lcp h\pi\pi) - M([pK^-\pi^+]_{\Lcp}) + M_{\rm PDG}(\Lcp),
\end{eqnarray}
where $h\pi\pi$ indicates $\Km\pip\pip$ ($\pim\pip\pim$) for \Xiccpp
(\Lb) candidates, 
$M(\Lcp h\pi\pi)$ is the invariant mass of the $\Xiccpp$ or $\Lb$
candidate, $M([pK^-\pi^+]_{\Lcp})$ is the invariant mass of the $\Lcp$
candidate, and $M_{\rm PDG}(\Lcp)$ is the known value of the \Lc mass~\cite{PDG2017}.
The distributions of the mass $m$ of selected
$\Lcp\Km\pip\pip$ and $\Lcp \pim \pip \pim$ candidates
are shown in
Fig.~\ref{fig:MassSys-Xicc}.
Unbinned extended maximum-likelihood fits to these distributions are
performed as in Ref.~\cite{LHCb-PAPER-2017-018}, with
the signal described by the sum of a Gaussian function and a double-sided Crystal Ball
function~\cite{Skwarnicki:1986xj}, and the background parameterised by 
a second-order Chebyshev polynomial.
The same fit models are used for both the $\Xiccpp$ and $\Lb$ samples,
but with different resolution parameters.
Signal yields of
$304 \pm 35$ \Xiccpp and
$ 3397 \pm 119$ \Lb decays are obtained.
The small decrease in the $\Xiccpp$ yield
compared with the value of $313 \pm 33$ reported in Ref.~\cite{LHCb-PAPER-2017-018}
is due to the two additional selection requirements described above.

\begin{figure}
\begin{center}
    \includegraphics*[width=.49\textwidth]{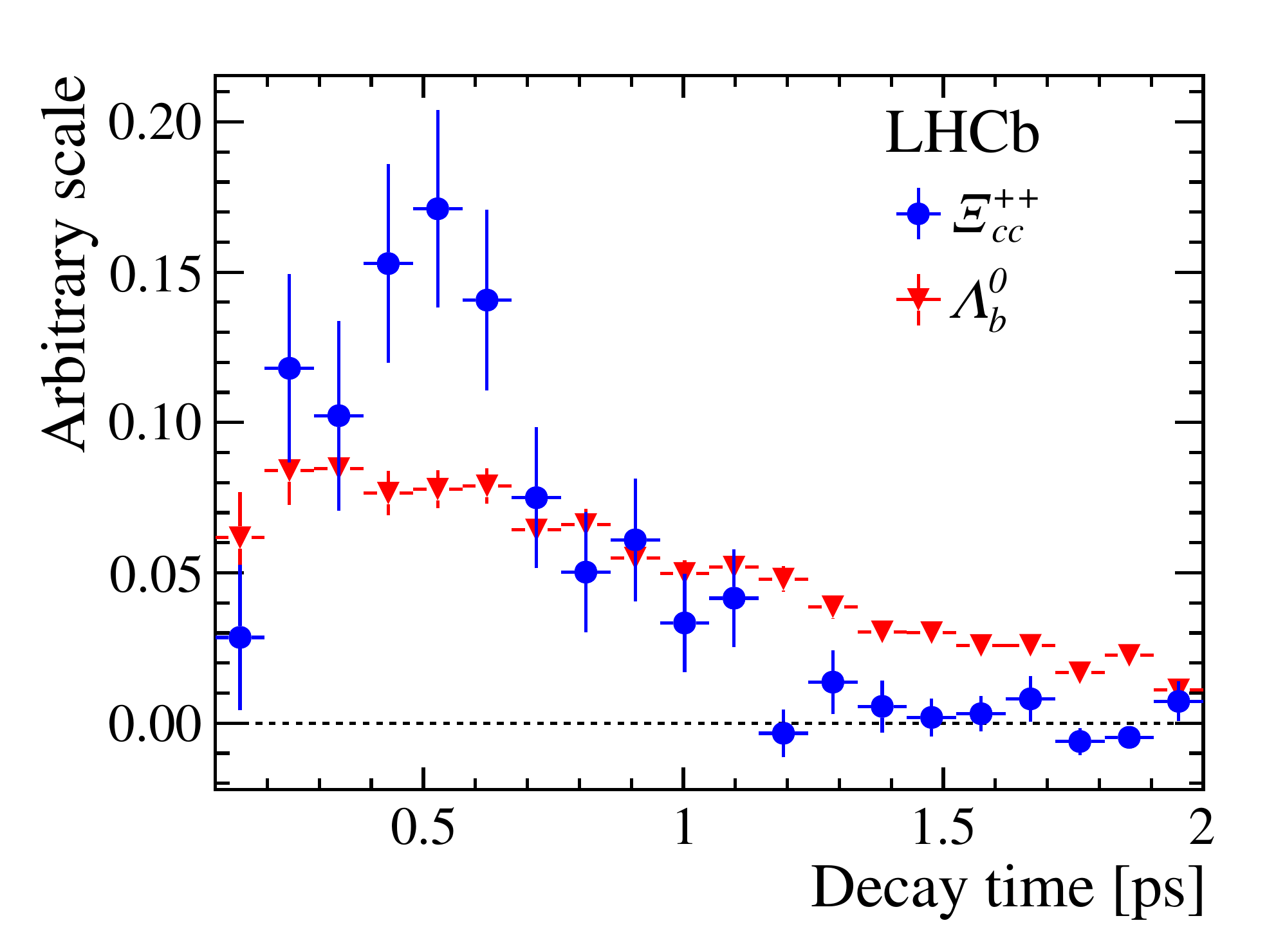}
\end{center}
\vspace*{-.7cm}
\caption{\small Background-subtracted decay-time distributions of
  (dots) $\Xiccpp \to \Lcp\Km\pip\pip$ 
 and (triangles) $\Lb \to \Lcp \pim \pip\pim$ candidates after the selection,
 not corrected for decay-time acceptance.
}
\label{fig:RawDecayTime}
\end{figure}

The decay time of \Xiccpp or \Lb candidates is computed
with a kinematic fit~\cite{Hulsbergen:2005pu} 
in which
the momentum vector of the candidate is required to be aligned 
with the line joining the production and decay vertices. 
The decay-time resolution, determined from simulation, is 63\fs (32\fs) for the \Xiccpp (\Lb) decay,
which is much less than the $\Xiccpp$ ($\Lb$) lifetime and 
has negligible dependence on the decay time within the current
precision.
The normalised decay-time distributions of the $\Xiccpp$ and $\Lb$ baryons 
are shown in Fig.~\ref{fig:RawDecayTime},
where the background contributions have been subtracted 
according to the fit results shown in Fig.~\ref{fig:MassSys-Xicc}
using the \sPlot technique~\cite{sWeights}.

\begin{figure}
\begin{center}
    \includegraphics*[width=.49\textwidth]{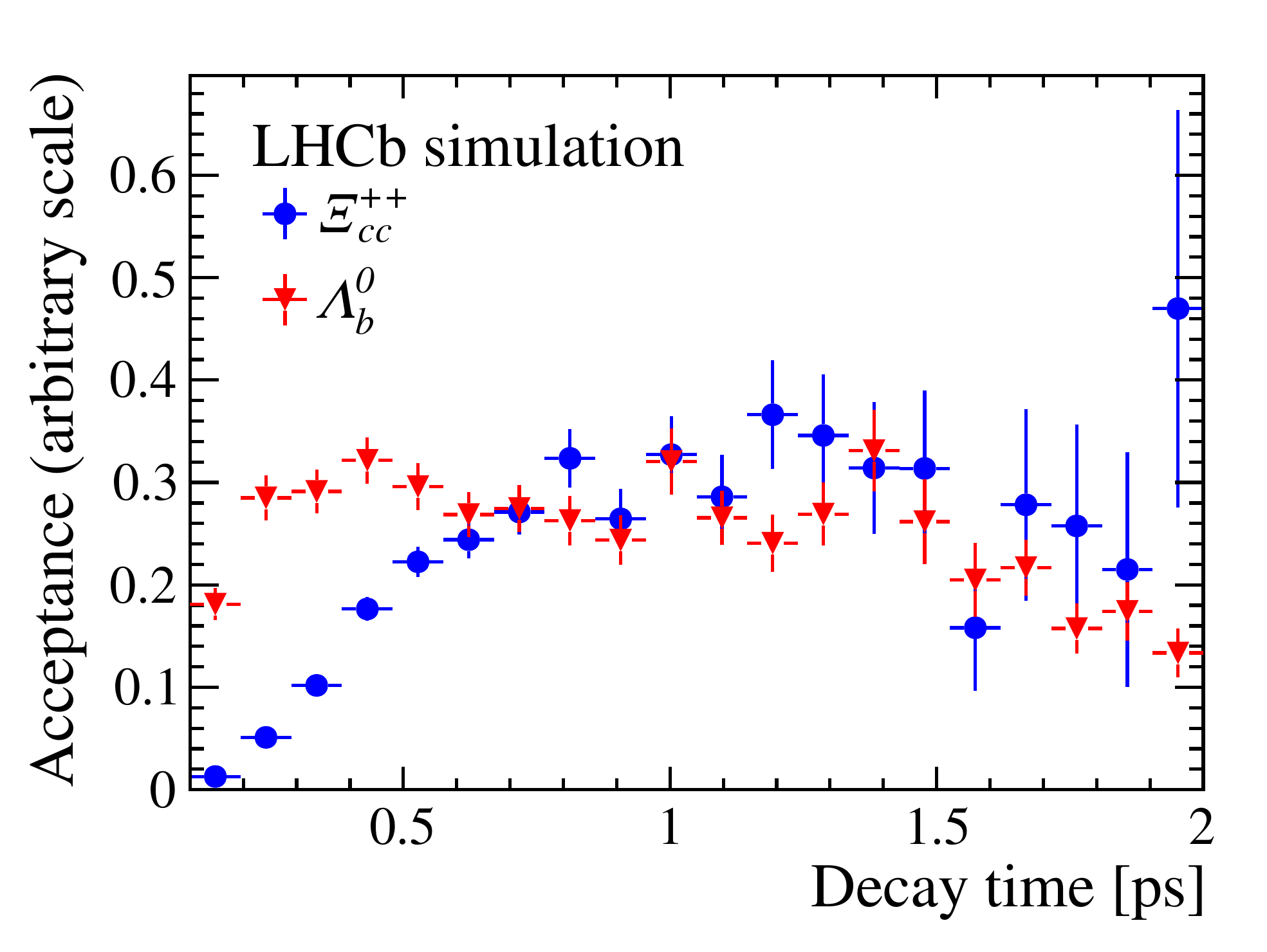}
\end{center}
\vspace*{-.7cm}
\caption{\small Decay-time acceptances
                for (dots) $\Xiccpp \to \Lcp\Km\pip\pip$  
                 and (triangles) $\mbox{\Lb \to \Lcp \pim \pip\pim}$ decays.
}
\label{fig:Acceptance}
\end{figure}

The decay-time acceptance is defined as the ratio between 
the reconstructed and the generated decay-time distributions, 
and is determined with samples of simulated events
containing \Xiccpp (\Lb) decays, in which the $\Xiccpp$ ($\Lb$)
lifetime is set to 0.333 ps (1.451 ps), as shown in
Fig.~\ref{fig:Acceptance}. 
This decay-time acceptance, which is described by a histogram in this analysis,
takes into account the reconstruction efficiency, 
as well as the bin migration effect  caused by the decay-time resolution. 
A potential bias in the relative decay-time acceptance due to the assumed lifetimes
is considered as a source of systematic uncertainty.  
The simulated $\Xiccpp$ and $\Lb$ decays are weighted to match their
observed transverse-momentum
distributions in data.  
The difference between the \Xiccpp or \Lb decay-time acceptances
is mainly due to the larger $\Lb$ mass, which results in higher
momentum of the decay products and larger opening angles in the decay.
An exponential function is fitted to the background-subtracted and  
acceptance-corrected decay-time distribution of $\Lb$ candidates,
and a lifetime of $1.474 \pm 0.077\ps$ is obtained, 
where the uncertainty is statistical only. 
This is consistent with the known value $1.470 \pm
0.010\ps$~\cite{PDG2017}, 
and validates that the detector simulation correctly reproduces the decay-time acceptance.

\begin{figure}
\begin{center}
    \includegraphics*[width=.49\textwidth]{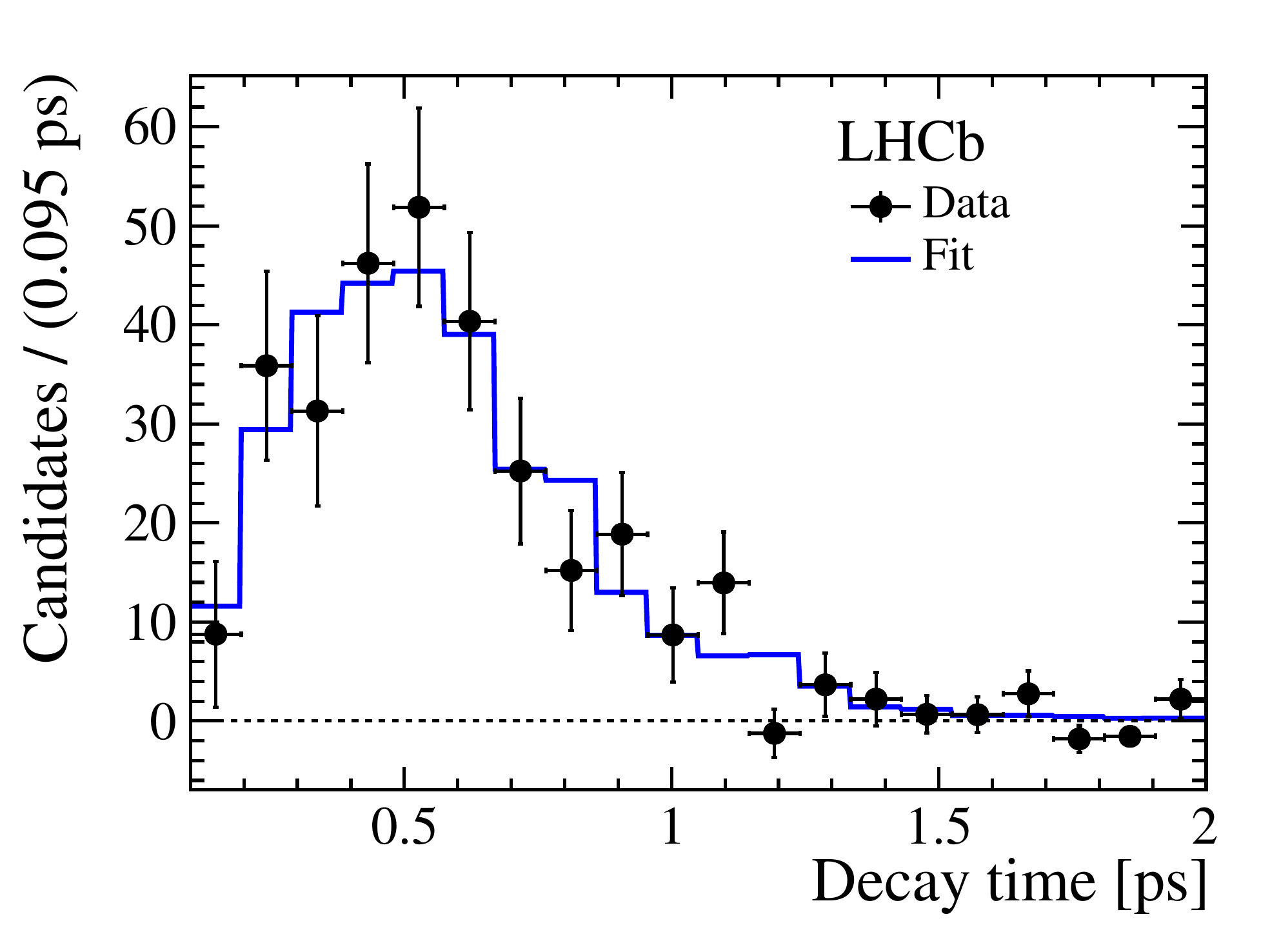}
\end{center}
\vspace*{-.7cm}
\caption{\small Background-subtracted decay-time distribution of selected 
$\Xiccpp \to\Lcp\Km\pip\pip$ candidates. 
The rate-averaged fit result 
across each decay-time bin is shown as the continuous line.}
\label{fig:DecayTime}
\end{figure}

The $\Xiccpp$ lifetime is measured by performing a weighted, unbinned maximum-likelihood fit~\cite{sFit}
to  the decay-time distribution of the selected  $\Xiccpp$ sample. Each candidate  is 
assigned a signal weight for background subtraction, which is computed using
its invariant mass $m$ as the discriminating variable following the \sPlot technique~\cite{sWeights}. 
The probability density function describing the decay-time distribution of the $\Xiccpp$  signal candidates, 
denoted by $f_{\Xiccpp}(t)$, is defined as 
\begin{equation}
f_{\Xiccpp}(t) = H_{\Lb}(t) \times 
\frac{\epsilon_{\Xiccpp}(t)}{\epsilon_{\Lb}(t)} \times 
\exp\left( \frac{t}{\SymbolLbTau} - \frac{t}{\SymbolXiccTau} \right),
\label{eq:Unbinnedfunction}
\end{equation}
where $H_{\Lb}(t)$ is the background-subtracted decay-time distribution
of the \Lb control channel,
$\epsilon_{\Xiccpp}(t)$ and $\epsilon_{\Lb}(t)$  are the decay-time acceptance distributions for
the \Xiccpp and \Lb decays,
and $\SymbolLbTau = 1.470 \pm 0.010$\ps is the known
value~\cite{PDG2017} of the $\Lb$ lifetime~\cite{LHCb-PAPER-2014-003}.
Here $H_{\Lb}(t)$, $\epsilon_{\Xiccpp}(t)$, and $\epsilon_{\Lb}(t)$
are the histograms shown in Figs.~\ref{fig:RawDecayTime} and~\ref{fig:Acceptance}.
The binning scheme is chosen 
to minimize the systematic uncertainty on the lifetime due to the finite bin width.
The background-subtracted  \Xiccpp decay-time distribution 
is shown in Fig.~\ref{fig:DecayTime} with the fit result superimposed.
The only free parameter of the fit is the \Xiccpp lifetime, 
which is measured to be
$\SymbolXiccTau = \XiccTauStat$. Here the uncertainties are statistical only,
and include contributions due to the limited sizes 
of the simulated samples ($0.007$\ps) and of the \Lb sample
($0.006$\ps). 
These contributions are estimated with a bootstrapping method~\cite{efron1979},
where candidates are randomly selected from the original simulated
or $\Lb$ samples to form statistically independent samples of pseudodata.
The standard deviations of the lifetime measurements obtained in these samples
are then taken as the corresponding statistical uncertainty.

Sources of systematic uncertainty on the \Xiccpp
lifetime are summarised in Table~\ref{tab:systematics}
and described below.
The effects of the choice of signal and background models
are studied by using alternative mass shapes, namely a sum of two Gaussian functions
for signal and an exponential function 
for background. The change in the measured lifetime, $0.005$\ps, is assigned 
as a systematic uncertainty.
In the baseline fit, the signal and background mass shapes are assumed to be independent 
of the decay time. The effect of this assumption is investigated by fitting 
the invariant-mass distribution of the \Xiccpp and \Lb samples 
in four independent intervals of decay time
and recalculating the signal weights based on these fit results. 
Using these weights in the fit, the \Xiccpp lifetime changes by $0.004$\ps,  
which is taken as the systematic uncertainty due to the correlation between the mass 
and decay time.
It is found that the measured lifetime
depends slightly upon the binning scheme. With the nominal binning,
a difference of $0.001$\ps with respect to the input lifetime is
measured,
which is taken as a systematic uncertainty. 

The kinematic distributions of the $\Xiccpp$ and $\Lb$ signals 
in the simulation are generally
found to be in good agreement with those in data. 
However, some differences
are observed in the output distribution of the multivariate
selector. To assess the impact of such differences, the simulation
is weighted to match this  output distribution  in data
 and the decay-time acceptance is recomputed.
The difference between the result from this procedure and the original one is $0.004$\ps,
which is assigned as the corresponding systematic uncertainty.
The simulated $\Xiccpp \to \Lcp\Km\pip\pip$ and $\Lb \to \Lcp\pim\pip\pim$
samples are generated assuming that the decay products are distributed 
uniformly across the available phase space. 
The possible effect of intermediate resonances
is evaluated by weighting the
simulated invariant mass distributions of the three hadrons,
\ie $M(\Km\pip\pip)$ for \Xiccpp and $M(\pim\pip\pim)$ for \Lb candidates,
to match the distributions seen in data.
The resulting difference in the measured lifetime,
$0.011$\ps, is assigned as a systematic uncertainty. 

The transverse-energy threshold in the calorimeter hardware trigger
varied during data taking, and
this variation is not fully described by the simulation.
To investigate the influence of this difference,
the hardware trigger requirement is applied to the data with a higher (uniform) threshold.
The measurement is repeated 
and the change in the measured lifetime, $0.002$\ps, is taken as a systematic uncertainty.
The input lifetime used in the simulation for the \Xiccpp baryon 
is $0.333$\ps. The simulated events are weighted to be distributed
according to the measured
lifetime
and the decay-time acceptance is recomputed.
The resulting difference in the measured lifetime, $0.002$\ps, 
is taken as a systematic uncertainty.
The \Lb lifetime is precisely known~\cite{PDG2017,LHCb-PAPER-2014-003}.
An alternative fit in which $\SymbolLbTau$ is allowed to vary within
its uncertainty leads to a change in the measured \Xiccpp lifetime of
less than $0.001$\ps, which is assigned as a systematic uncertainty. 

Other systematic effects, including 
the threshold applied to the multivariate selector,
the decay-time resolution, 
and the uncertainty on the length scale of the vertex detector,
are studied and found to be negligible;
no systematic uncertainties are assigned for these effects.
As further checks, the measured lifetime is 
compared between subsets of the data, 
including $\Xiccpp$ versus $\Xiresbar_{cc}^{--}$, opposite LHCb magnet polarities, and
different numbers of primary vertices, and is found to be stable.
A separate measurement carried out with an alternative method,
in which both the \Xiccpp and \Lb decay-time distributions 
are binned, gives a consistent result. 
All sources of systematic uncertainty, listed in
Table~\ref{tab:systematics}, are added in quadrature, 
and the total systematic uncertainty on the 
measured \Xiccpp lifetime is found to be $0.014$\ps. 

\begin{table}
\begin{center}
\caption{Summary of systematic uncertainties.}
\begin{tabular}{l c}
\hline\hline
Source  & Uncertainty (ps) \\ 
\hline
Signal and background mass models & $0.005$ \\
Correlation of mass and decay-time & $0.004$ \\
Binning  & $0.001$ \\
Data-simulation differences & $0.004$ \\
Resonant structure of decays &  $0.011$ \\
Hardware trigger threshold & $0.002$ \\
Simulated \Xiccpp lifetime & $0.002$ \\
\Lb lifetime uncertainty & $0.001$ \\ \hline
Sum in quadrature & $0.014$ \\
\hline
\hline
\end{tabular}
\label{tab:systematics}
\end{center}
\end{table}

In summary, the $\Xiccpp$ lifetime is measured 
using a data sample corresponding to
an integrated luminosity of 1.7\invfb, collected 
by the LHCb experiment in $pp$ collisions at a centre-of-mass energy of 13\tev,
and is found to be
\begin{equation}
\SymbolXiccTau = \XiccTau.\nonumber
\end{equation}
This is the first measurement of the \Xiccpp lifetime,
which establishes the weakly decaying nature of the
recently discovered $\Xiccpp$ state.
The result favours smaller values
in the range of the theoretical predictions~\cite{
Fleck:1989mb,
guberina1999inclusive,
Kiselev:1998sy,
Likhoded:1999yv,
Onishchenko:2000yp,
Anikeev:2001rk,
Kiselev:2001fw,
hsi2008lifetime,
Karliner:2014gca,
Berezhnoy:2016wix}. 
If the lifetime of the isospin partner state \Xiccp is shorter
by a factor of 3 to 4 as predicted~\cite{
Fleck:1989mb,
guberina1999inclusive,
Kiselev:1998sy,
Likhoded:1999yv,
Onishchenko:2000yp,
Anikeev:2001rk,
Kiselev:2001fw,
hsi2008lifetime,
Karliner:2014gca,
Berezhnoy:2016wix}, 
it would be roughly 60--90\fs.
This provides important information to guide the search for the
$\Xiccp$ state at the Large Hadron Collider.

\section*{Acknowledgements}
\noindent 
We thank Chao-Hsi Chang, Cai-Dian L\"u, Wei Wang, Xing-Gang Wu, and
Fu-Sheng Yu for frequent and interesting discussions on the production
and decays of double-heavy-flavor baryons.
We express our gratitude to our colleagues in the CERN
accelerator departments for the excellent performance of the LHC. We
thank the technical and administrative staff at the LHCb
institutes. We acknowledge support from CERN and from the national
agencies: CAPES, CNPq, FAPERJ and FINEP (Brazil); MOST and NSFC
(China); CNRS/IN2P3 (France); BMBF, DFG and MPG (Germany); INFN
(Italy); NWO (Netherlands); MNiSW and NCN (Poland); MEN/IFA
(Romania); MinES and FASO (Russia); MinECo (Spain); SNSF and SER
(Switzerland); NASU (Ukraine); STFC (United Kingdom); NSF (USA).  We
acknowledge the computing resources that are provided by CERN, IN2P3
(France), KIT and DESY (Germany), INFN (Italy), SURF (Netherlands),
PIC (Spain), GridPP (United Kingdom), RRCKI and Yandex
LLC (Russia), CSCS (Switzerland), IFIN-HH (Romania), CBPF (Brazil),
PL-GRID (Poland) and OSC (USA). We are indebted to the communities
behind the multiple open-source software packages on which we depend.
Individual groups or members have received support from AvH Foundation
(Germany), EPLANET, Marie Sk\l{}odowska-Curie Actions and ERC
(European Union), ANR, Labex P2IO and OCEVU, and R\'{e}gion
Auvergne-Rh\^{o}ne-Alpes (France), Key Research Program of Frontier
Sciences of CAS, CAS PIFI, and the Thousand Talents Program (China),
RFBR, RSF and Yandex LLC (Russia), GVA, XuntaGal and GENCAT (Spain),
Herchel Smith Fund, the Royal Society, the English-Speaking Union and
the Leverhulme Trust (United Kingdom).

\addcontentsline{toc}{section}{References}
\setboolean{inbibliography}{true}
\bibliographystyle{LHCb}
\bibliography{main}

\newpage
\centerline{\large\bf LHCb collaboration}
\begin{flushleft}
\small
R.~Aaij$^{27}$,
B.~Adeva$^{41}$,
M.~Adinolfi$^{48}$,
C.A.~Aidala$^{73}$,
Z.~Ajaltouni$^{5}$,
S.~Akar$^{59}$,
P.~Albicocco$^{18}$,
J.~Albrecht$^{10}$,
F.~Alessio$^{42}$,
M.~Alexander$^{53}$
% ...

{\footnotesize \it
% ...
$ ^{5}$Clermont Universit{\'e}, Universit{\'e} Blaise Pascal, CNRS/IN2P3, LPC, Clermont-Ferrand, France\\
% ...
$ ^{10}$Fakult{\"a}t Physik, Technische Universit{\"a}t Dortmund, Dortmund, Germany\\
% ...
$ ^{18}$INFN Laboratori Nazionali di Frascati, Frascati, Italy\\
% ...
$ ^{27}$Nikhef National Institute for Subatomic Physics, Amsterdam, Netherlands\\
% ...
$ ^{41}$Instituto Galego de F{\'\i}sica de Altas Enerx{\'\i}as (IGFAE), Universidade de Santiago de Compostela, Santiago de Compostela, Spain\\
$ ^{42}$European Organization for Nuclear Research (CERN), Geneva, Switzerland\\
% ...
$ ^{48}$H.H. Wills Physics Laboratory, University of Bristol, Bristol, United Kingdom\\
% ...
$ ^{53}$School of Physics and Astronomy, University of Glasgow, Glasgow, United Kingdom\\
% ...
$ ^{59}$University of Cincinnati, Cincinnati, OH, United States\\
% ...
$ ^{61}$Syracuse University, Syracuse, NY, United States\\
% ...
$ ^{73}$University of Michigan, Ann Arbor, United States, associated to $^{61}$\\
% ...
}
\end{flushleft}

\end{document}